# Genuine Ohmic van der Waals contact between indium and MoS₂

Bum-Kyu Kim[1†], Dong-Hwan Choi[1,2†], Tae-Hyung Kim[3†], Hanul Kim[2], Kenji Watanabe[4], Takashi Taniguchi[4], Heesuk Rho[2], Yong-Hoon Kim[3*], Ju-Jin Kim[2*], and Myung-Ho Bae[1,5*]

[1]Korea Research Institute of Standards and Science, Daejeon 34113, Republic of Korea

[2]Department of Physics, Research Institute of Physics and Chemistry, Chonbuk National University, Jeonju 54896, Republic of Korea

[3]School of Electrical Engineering, Korea Advanced Institute of Science and Technology, 291 Daehak-ro, Yuseong-gu, Daejeon 34141, Republic of Korea

[4]National Institute for Materials Science, 1-1 Namiki, Tsukuba 305-0044, Japan

[5]Department of Nano Science, University of Science and Technology, Daejeon, 34113, Republic of Korea

[†]These authors contributed equally to this work.

*e-mail: y.h.kim@kaist.ac.kr; jujinkim@chonbuk.ac.kr; mhbae@kriss.re.kr

**Abstract**

**The formation of an ideal van der Waals (vdW) contacts at metal/transition-metal dichalcogenide (TMDC) interfaces is a critical step for the development of high-performance and energy-efficient electronic and optoelectronic applications based on the two-dimensional (2D) semiconductors. In overcoming the key challenges of the conventional metal deposition process that leads to an uncontrollable Schottky barrier height and high contact resistance, notable advances were recently made by transferring atomically flat metal thin films[1] or thermally evaporating indium/gold alloy[2]. However, the realization of an ideal vdW contact between an elemental metal and TMDC through the evaporation process is yet to be demonstrated,**

**and particularly the evidence of an Ohmic contact between three-dimensional metallic electrodes and TMDCs is still unavailable. Herein, we report the fabrication of atomically clean metal/TMDC contacts by evaporating metals at a relatively low thermal energy and subsequently cooling the substrate holder down to ~100 K by liquid nitrogen, achieving for the indium (In)/molybdenum disulfide (MoS$_2$) case an accumulation-type Ohmic contact with a metal-induced electron doping density of ~$10^{12}$ cm$^{-2}$. We find that the transport at the In/MoS$_2$ contact is dominated by the field-emission mechanism over a wide temperature range from 2.4 to 300 K, and the contact resistance reaches ~ 600 Ω μm and ~ 1,000 Ω μm at cryogenic temperatures for the few-layer and monolayer MoS$_2$ cases, respectively. Based on first-principles calculations, we find that the nature of the ideal In/MoS$_2$ vdW contact is characterized by the formation of in-gap states within TMDC together with the abrupt and rigid shift of the TMDC band. Demonstrating the genuine Ohmic vdW contacts through the evaporation process, this work paves a practically available path for fabricating high-quality metal/2D semiconductor contacts for next-generation electronic and optoelectronic applications.**

Layered semiconducting transition-metal dichalcogenides (TMDCs) such as MoS$_2$, WSe$_2$ and MoTe$_2$ have been extensively studied for the future development of low-power and high-performance electronics and optoelectronics[3,4,5]. In realizing two-dimensional (2D) electronic devices based on TMDCs, because each TMDC atomic layer in the bulk form is coupled with neighboring layers through van der Waals (vdW) interactions, mono- or few-layer TMDC flakes can be deposited onto a substrate via a mechanical exfoliation method[3]. Meanwhile, establishing a reliable

Ohmic contact across the Schottky barriers between metals and TMDCs remains a critical challenge[6,7]. For instance, in reducing the Schottky barrier height for TMDCs, efforts to identify metals with appropriate work functions $\Phi_{metal}$ (e.g. $\Phi_{Sc} \approx 3.5$ eV for Sc and $\Phi_{Ti} \approx 4.3$ eV for Ti) based on the affinity of TMDC (e.g. monolayer $MoS_2$: $\chi_{1L\,MoS_2} \approx 4$ eV, multilayer $MoS_2$: $\chi_{ML\,MoS_2} \approx$ 4.3 eV) have not been effective because of the strong Fermi-level pinning (FLP) effect[8,9]. While various approaches have been explored to overcome this problem, including molecular doping[10], tunnel-barrier insertion[11,12], fabrication of graphene contacts[13,14], and phase changes[15], a recent study has shown that the formation of an ideal metal-TMDC vdW contact through the transfer of atomically flat metal thin films significantly improves the contact properties.[1] Here, it was important to recognize that the conventional thermal evaporation process of metals typically introduces crystalline defects in TMDCs and leads to an uncontrollable Schottky barrier height (or FLP) and high contact resistance[9,1]. More recently, another breakthrough was made by introducing the thermal-evaporation process of In alloy with Au, leading to clean vdW-type contacts for single- and few-layer $MoS_2$ and achieving the lowest contact resistance between three-dimensional (3D) metals and 2D semiconductors[2]. However, question still remains whether it is also possible to form an ideal vdW contact for 2D semiconductors with a single elemental metal through the evaporation process. Furthermore, if the low contact resistance through the vdW contact can be achieved, one needs to elucidate the origins, i.e., what is the nature of the charge transport and band alignment at the interface and whether the transport across the contacts is thermionic or field emission.

Here, we report the Ohmic vdW contact formed between pure In and $MoS_2$, and characterize the In/$MoS_2$ interface by the transmission electron microscope (TEM) measurement, transport measurements at varying temperature ($T$) and carrier density ($n_e$), and first-principles calculations. A key to this achievement was the relatively low-evaporation $T$ of In at ~530 °C with the extremely low $T$ of the substrate holder at ~100 K in the In deposition process, which allows the preparation of an atomically clean In/MoS$_2$ interface. If an ideal vdW-type contact is achieved, the relatively low work function of In ($\Phi_{In} \approx 4.1$ eV) should result in a good contact resistance for MoS$_2$. For both single- and few-layer MoS$_2$ devices prepared through our approach, we indeed find that the contact resistance decreases with decreasing temperature, reaching to the record-level 1 k$\Omega$ μm and 0.6 k$\Omega$ μm, respectively. This behavior indicates the Ohmic character of the In/MoS$_2$ contact or the field emission is the dominant charge transport mechanism at the In/MoS$_2$ contact. Carrying out first-principles calculations for realistic In/MoS$_2$ contact models, we find that the atomistic origins of the high-quality Ohmic vdW contact can be traced to the weak In/MoS$_2$ interactions and yet the formation of in-gap states within the contact-region MoS$_2$ to support the interfacial charge transport.

## Preparation and characterization of van-der-Waals In/MoS$_2$ interface

We fabricated MoS$_2$ field-effect transistors (FETs) on hBN flakes, where the hBN flakes were deposited onto a 300-nm-thick SiO$_2$/Si substrate by mechanical exfoliation. We then transferred a few-layer MoS$_2$ flake (HQ-graphene, Inc.) onto a selected hBN flake[13,16]. For the electrical measurements, we deposited 100 nm-thick In electrodes across the MoS$_2$ channel, where the substrate holder was kept at ~100 K by flowing liquid nitrogen through it (see Fig. 1a). The substrate cooling process leads to an important result; a uniform surface morphology of indium film is achieved, which contrasts strongly with the usage of a room-temperature holder that produces a segregated granular film for at least up to ~ 70 nm thickness, as reported in our previous work[17]. The upper panel of Fig. 1b shows a cross-sectional transmission electron microscopy (TEM) image of the In/few-layered MoS$_2$ junction of a separately prepared sample, which clearly shows an atomically separated interface between In and MoS$_2$ layers without any metal invasion into the MoS$_2$ layers. Whereas crystal-lattice disorders that cause defect-induced gap states and FLP typically occur during the high-temperature deposition process of evaporated metal atoms with high thermal energy[9,1], the In deposited at a relatively low thermal energy could apparently provide a clean vdW interface without disorder or defect. For instance, whereas the evaporation temperature of Au at $10^{-7}$ Torr is ~860 °C, only ~530 °C is required for the evaporation of In at the same pressure. For comparison, we also prepared an Au/few-layered MoS$_2$ junction, where the substrate holder was also kept at ~100 K during Au deposition. The lower panel of Fig. 1b shows a TEM image of the junction having atomic defects at the first and second layers from the top of MoS$_2$, due to the invasion of Au atoms during deposition process.

## Basic electrical properties of a few-layer MoS$_2$ field-effect transistor

Figure 1c shows a photograph of a MoS$_2$ FET on a 22 nm-thick hBN flake. The number of MoS$_2$ layers was estimated to be $n = 6$ (see Supplementary Fig. S1). The multiple electrodes for the six-layer (6L) MoS$_2$ flake with different intervals between two neighboring electrodes were designed to measure the contact resistance via the transfer-length method (TLM)[15], i.e., four FETs with different channel lengths ($L_1 = 0.5$ μm, $L_2 = 1$ μm, $L_3 = 1.5$ μm, and $L_4 = 2$ μm from the left channel in the region indicated by a dashed box). Here, the widths ($W$) of all channels were nearly identical at 2 μm. Figure 1d shows the two-probe conductance as a function of the back-gate voltage ($G$–$V_G$) of the $L_2$-FET with a source–drain voltage ($V_{SD}$) of 30 mV at various temperatures. The conductance decreased for negatively decreasing $V_G$ and reached zero near $V_G \approx 0$ V throughout the investigated temperature range, which indicates that the carriers are electrons. The two-probe conductance increased with decreasing $T$ at a given $V_G$ for $V_G > 10$ V, i.e., the device exhibited a metallic behavior. However, the opposite behavior was observed near a depletion region of $0 < V_G < 10$ V, representing an insulating character. These behaviors are consistent with the current–voltage ($I$–$V_{SD}$) curves for various $V_G$ values at $T = 2.5$ K in Supplementary Fig. S2a. For instance, the $I$–$V_{SD}$ curves for $V_G > 10$ V and $0 < V_G < 10$ V show linear and nonlinear characteristics, respectively. The four-probe measurements for the $L_2$ channel also showed a similar $V_G$ value for the metal–insulator crossover location (see Supplementary Fig. S2b and S2c). This result indicates that the crossover behavior is mainly determined by the transport in the $MoS_2$ channel, not in the contact part. In the field-effect mobility ($\mu$) as a function of $T$ (see Supplementary Fig. S3), the mobility showed the temperature dependence of $T^{-2.2}$ at $T > 100$ K with $\mu \approx 50$ cm$^2$V$^{-1}$s$^{-1}$ at room temperature. This behavior is similar to the bulk $MoS_2$ ($\mu \propto T^{-2.5}$) with optical phonon scattering as the dominant scattering mechanism[18,13]. When $T < 20$ K, $\mu$ became saturated at 3200 cm$^2$ V$^{-1}$ s$^{-1}$ with decreasing temperature for the four-probe measurement. The saturation behavior in the low-$T$ regime has been known to occur when the impurity scattering plays a dominant role and the phonon effect is suppressed[19].

## Contact resistance at In/MoS₂ contacts

On the basis of the TLM measurements (see Supplementary Fig. S4) with multiple channels (see Fig. 1c), we extracted the contact resistance ($R_cW$) as a function of $n_e$ of a 6L-$MoS_2$ device at representative temperatures; the results are shown in Fig. 2a (scattered solid squares). Here, $n_e$ was estimated from the relation $n_e = (e\mu R_{sh})^{-1}$, where the sheet resistance $R_{sh}$ and $\mu$ were obtained from the four-probe measurements. We note that the obtained contact resistance includes a serial resistance of In electrodes. At a given temperature, $R_cW$ decreased with increasing $n_e$. The contact resistance is given by[6,7]

$$R_c W\,(n_e, T)\, =\, \sqrt{R_{sh}(n_e, T)\rho_c(n_e, T)},\tag{1}$$

which is only valid for $L_c \gg L_T$. Here, $L_c$ and $L_T$ $(= \sqrt{\rho_c / R_{sh}})$ are respectively the contact length and transfer length, the latter of which represents the average distance that charge carriers flow in a semiconductor beneath the contact before they completely transport to the electrode. Figure S4 shows that our device satisfied this condition with $L_c \approx 1$ μm and $L_T \approx 0.1$ μm. Equation (1) implies that $R_c W$ decreases with increasing $n_e$ because both $R_{sh}$ and $\rho_c$ generally decrease with increasing $n_e$.

At a fixed $n_e$, whereas the thermionic emission charge transport mechanism across the Schottky barrier predicts that $R_c W$ increases with decreasing $T$ because of suppression of the thermionic emission[20], $R_c W$ in our measurements decreased with decreasing $T$ as shown in Fig. 2a. Figure 2b shows $R_c W$ as a function of $T$ at $n_e = 3.4 \times 10^{12}$ cm$^{-2}$. The contact resistance decreased from 2.3 to 0.6 kΩ μm when $T$ was decreased from room temperature to 2.4 K. This behavior has been reported in several previous works such as graphene/MoS$_2$[13] and Pd/graphene contacts[21]; it is considered as an evidence for the non-dominant role of thermionic emission for the contact transport mechanism. Figure 2b also shows $R_{sh}$ as a function of $T$ at $n_e = 3.4 \times 10^{12}$ cm$^{-2}$, where $R_{sh}$ decreased with decreasing $T$ because the phonon scattering is reduced.

**Mechanism of transport at In/MoS$_2$ contacts**

We next examined which component between $R_{sh}$ and $\rho_c$ predominantly determines the contact resistance of the In/MoS$_2$ vdW contact. We included other reports in Fig. 2a for comparison; graphene(Gr)/four-layer (4L)-MoS$_2$ ($n = 4$) contact,[13] Au/4L-MoS$_2$ [20] and Au/In/few-layer MoS$_2$ [2]. In the case of Gr/4L-MoS$_2$, the graphene functions as a work-function-controllable contact material, which leads to a lower contact resistance, i.e., $R_cW \approx 1$ k$\Omega$ μm at $n_e > 4 \times 10^{12}$ cm$^{-2}$ and $T = 12$ K (see the red curve in Fig. 2a). Although both the Gr-and In-contact MoS$_2$ devices gave a similar minimum $R_cW$ at cryogenic temperatures, we conclude that the transport mechanisms at the contacts differ from each other. In the In/MoS$_2$ contact case, $R_cW$ decreased with decreasing $T$ for the examined $n_e$ range, representing the field emission (or tunneling) for all examined $T$ and $n_e$ ranges. However, the $R_cW$–$n_e$ curves obtained at $T = 12$ and 250 K for the Gr/4L-MoS$_2$ device suggest that the left and right sides with respect to $n_e \approx 2 \times 10^{12}$ cm$^{-2}$ followed the thermionic and field emissions at the contact, respectively. At $T = 12$ K for the Gr/4L-MoS$_2$ device, although the $R_cW$ of ~1 k$\Omega$ μm was relatively insensitive to the variation of $n_e$ in the range from $4 \times 10^{12}$ to $7 \times 10^{12}$ cm$^{-2}$, it rapidly changed from 1 k$\Omega$ μm to 6 k$\Omega$ μm when $n_e$ decreased from $3 \times 10^{12}$ cm$^{-2}$ to $1.5 \times 10^{12}$ cm$^{-2}$. In the case of the Au/4L-MoS$_2$ contact, $R_cW$ was increased with decreasing temperature, representing the thermionic emission. For the Ohmic contact case of the In/6L-MoS$_2$ device, on the other hand, the slopes in the $R_cW$-$n_e$ curves were nearly unchanged for varying temperature and $R_cW$ at a given temperature was also relatively insensitive to $n_e$. In this view, the slope of the Au/In/few layer-MoS$_2$ contact was similar to that for the thermal emission. In our In/6L-MoS$_2$ device (see Fig. 3a), $R_{sh}$ varies in the range from 1 to 80 k$\Omega$ when $\rho_c$ only varies from $5 \times 10^{-6}$ to $5 \times 10^{-7}$ $\Omega$ cm$^2$, as shown by two dashed lines, for $R_cW$ changing from 0.6 to ~3 k$\Omega$ μm. This result indicates that $R_{sh}$ plays a dominant role in determining the $R_cW$ in the field-emission region.

We also obtained $R_cW$ for the single-layer (1L)-MoS$_2$ device (see supplementary Fig. S6 for the thickness profile). In Fig. 3b, the $R_cW$ values were extracted via the TLM with three channels ($L = 0.5, 0.9, 1.4$ μm) as shown in the inset of Fig. 3b. For three $V_{G-th}$ conditions of 35, 40, and 45 V, $R_cW$ decreased with decreasing $T$ in the range $250 \geq T \geq 100$ K, representing the field emission. Here, $V_{G-th} = V_G$-$V_{th}$ and $V_{th}$ is the threshold voltage. At $V_{G-th} = 45$ V, $R_cW$ reached ~1 k$\Omega$ μm at $T$

= 100 K as the minimum value obtained from the 1L-MoS$_2$ device. This value is similar to that obtained from the 6L-MoS$_2$ device at a similar $T$ range (see Fig. 2). Interestingly, $R_c W$ increased with decreasing $T$ for $T < 100$ K under all $V_{G\text{-th}}$ conditions. In this region, the MoS$_2$ channel also exhibited an insulating behavior in $G$-$V_g$ curves for various temperatures of Fig. 3b for $V_G < 60$ V and $T < 100$ K. Thus, it indicates that the increase of $R_{sh}$ with decreasing $T$ in the insulating phase plays a dominant role in determining the contact resistance at $T < 100$ K. This implies that the manipulation of the metal-insulator crossover gate voltage could be crucial to get a better contact property in a single-layer MoS$_2$ device.

In Fig. 3a we compared the lowest achievable contact resistance as a function of $R_{sh}$ from previous reports with ours. For the In/MoS$_2$ device, the contact resistance decreased with decreasing $R_{sh}$ and reached ~0.6 kΩ μm as the lowest value at $R_{sh} \sim 1$ kΩ and $T = 2.4$ K. At $T = 300$ K, on the other hand, the Au/In/few layer-MoS$_2$ device showed the lowest value of ~0.6 kΩ μm at $R_{sh} \sim 60$ kΩ. Importantly, our In/MoS$_2$ contact provided 1-2 kΩ μm at $R_{sh} \sim 50$ kΩ as the lowest value for single-layer MoS$_2$ devices, while the Au/In/few layer-MoS$_2$ contact showed ~3 kΩ μm at $R_{sh} \sim 40$ kΩ.

## Accumulation-type contacts at In/MoS$_2$ interface

We note that both Fig. 1d and Fig. 3b show that the threshold $V_G$ for few-layer and monolayer MoS$_2$ is located at $V_G \geq 0$ V, which indicates that the Fermi energy $E_F$ of In is located near the conduction-band minimum (CBM) position of MoS$_2$. Since the electron affinity of MoS$_2$ ($\chi_{MoS_2}$) ranging from 4 eV (mono-layer MoS$_2$) to 4.3 eV (multilayer MoS$_2$) is comparable to the work function of In ($\Phi_{In}$) of ~ 4.1 eV, the vacuum (Schottky-Mott limit) band alignment at the interface can be expected to be an accumulation type[22]. To experimentally confirm the electron accumulation at an In-contacted region in $MoS_2$, we applied the Raman spectroscopy to measure the pristine and In-covered $MoS_2$ regions. It is known that the $A_{1g}$ phonon peak of $MoS_2$ exhibit a red shift and its width broadens upon electron doping[23,24]. Figures 4a and 4b show the optical images of respectively 1L- and bilayer (2L)-$MoS_2$ (indicated by dashed black lines) prepared on $SiO_2$ and partially covered with 5 nm-thick In (indicated by dashed white lines). Figures 4c and 4d show the $A_{1g}$ energy maps for the 1L- and 2L-$MoS_2$, respectively (see also Supplementary Fig. S7 for representative Raman spectra of 1L-$MoS_2$). The In-covered region shows a relatively lower energy than the non-covered region, i.e., $\Delta\omega \approx -0.3$ and $-1$ cm$^{-1}$ for the 1L- and 2L-$MoS_2$, respectively. These red shifts of the $A_{1g}$ energy reflect the electron doping of the $MoS_2$ under the In metal. Chakraborty et. al.[23] reported that the $A_{1g}$ mode softens with doping at a rate of ~0.2 cm$^{-1}$ per $10^{12}$ cm$^{-2}$ for 1L-$MoS_2$, indicating that the 1L-$MoS_2$ region covered by In was doped by electrons at a density of ~1.5×$10^{12}$ cm$^{-2}$. The full-width at half maximum, $\Gamma$, in Fig. 4e and 4f also shows consistent results. For instance, the In-covered region shows a relatively broader $\Gamma$ than the non-covered region for both 1L- and 2L-$MoS_2$, implying electron doping. The electron doping at the In/$MoS_2$ contact induces an accumulation-type contact, indicating the feasibility of field emission at the In/$MoS_2$ contact.

**Atomistic origin of low contact resistance at In/MoS$_2$ interface**

To clarify the special characters of indium in forming a vdW contact with $MoS_2$, we carried out comparative density functional theory (DFT) calculations for the vertical In/$MoS_2$ and Au/$MoS_2$ interface models shown in Fig. 5a. Shown together are the real-space charge density differences at the metal/$MoS_2$ interfaces calculated from the fully geometry-optimized models according to

$$\Delta\rho = \rho_{metal/MoS_2} - \left(\rho_{MoS_2} + \rho_{metal}\right). \tag{2}$$

We observe that, compared with the Au case, In induces much less charge transfer and concomitantly less structural distortions on $MoS_2$ (also see Supplementary Fig. S7). In addition, the direction of charge transfer in the $In/MoS_2$ contact was different from that in the $Au/MoS_2$ counterpart, and quantitatively the amounts of charge transfer from metal to $MoS_2$ were $+4.1 \times 10^{12}$ electrons/cm$^{-2}$ and $-2.5 \times 10^{13}$ electrons/cm$^{-2}$ for the In and Au cases, respectively. This indicates that In exhibits a relatively weak interactions with $MoS_2$ compared with Au. Moreover, the doping level calculated from the $In/MoS_2$ interface model is in good quantitative agreement with that estimated from Raman measurements, confirming the accuracy of our computational scheme (see Supplementary Fig. S8 for more details including additional consistency checks).

We next analyze the electronic structures of the $In/MoS_2$ interface in two steps. First, for the vertical $In/MoS_2$ model adopted above that represents the deep contact region (see Fig. 5b top panel), we calculated the projected density of states (DOS) and monitored how the $In-5p_z$ states evolve into the top $S-3p$, $Mo-4d$, and bottom $S-3p$ states (Fig. 5b bottom panels). We note that a noticeable amount of metal-induced gap states (MIGS) appear in the In-contacting top S layer as well as the Mo layer (indicated by blue arrows). Particularly, around the $MoS_2$ conduction band minimum (CBM) region located around the In Fermi level $E_F$ (indicated by a red arrow), we observe ample $Mo-4d$ states, which identifies the origin of the experimentally observed Ohmic transport behavior.

Because the vertical interface model is not sufficient to capture the development of the electrode-region $MoS_2$ electronic structure into that of the channel-region $MoS_2$, we next adopted a junction model shown in Fig. 5c top panel (see also Supplementary Fig. S9a) and carried out DFT-based matrix Green's function (MGF) calculations[25,26]. From the spatially-resolved DOS shown in Fig. 5c bottom panel, we immediately note that the above-discussed MIGS formed within the In-covered region (A, red upward triangle) propagates about ~ 1 nm into the channel region (B, blue upward triangle) before the intrinsic semiconducting channel region is recovered (C, green upward triangle). Importantly, taking the intrinsic $MoS_2$ channel region C as the reference, we can clarify the nature of the $MoS_2$ states around $E_F$ within the In-covered region A (Fig. 5b) as follows (see also Supplementary Information Fig. S9b): Forming an ideal FLP-free In/$MoS_2$ vdW contact, the respective vacuum-level electronic structures of In and $MoS_2$ (Fig. 5d; $\Phi_{In} \approx 4.1$ eV and $\chi_{1L\ MoS_2} \approx 4.2$ eV) hint the possibility of achieving an Ohmic behavior. However, the In-to-$MoS_2$ electron transfer should cause a band upshift, which then places the CBM of the intrinsic C-region $MoS_2$ at ~ 0.24 eV above $E_F$. The pure electrostatic argument can then predict the development of $MoS_2$ band from In to $MoS_2$ region A to $MoS_2$ region B in terms of the band bending picture such as given in Fig. 5e or Fig. 5f[6,7]. While the presence of $MoS_2$ states around $E_F$ in Fig. 5c might appear supporting this picture, a close inspection in fact presents another viewpoint as schematically summarized in Fig. 5g. Here, as is evident from the fact that the valence band maximum edge in region A does not bend upward into regions B and C in Fig. 5c, the upshift of $MoS_2$ band to the level expected in the intrinsic $MoS_2$ region C (+ 0.24 eV) should have abruptly occurred across the In-$MoS_2$ vdW gap. Then, the A-region $MoS_2$ states around $E_F$ in Fig. 5c, which were identified as the source of the Ohmic In/$MoS_2$ vdW contact, can be interpreted as the below-CBM MIGS rather than the bent-down CBM states, as depicted by thick red lines in Fig. 5g.

## Conclusion

In summary, we realized for the first time an Ohmic vdW contact between the 3D elemental metal In and 2D semiconductor $MoS_2$ by applying the low thermal-energy In deposition process while keeping the substrate holder at ~100 K. For the single- and few-layer MoS$_2$ devices, the contact resistance decreased with decreasing temperatures for $100 \leq T \leq 300$ K, indicating the field-emission mechanism for the Ohmic contact transport. The contact resistance was sensitive to the change of sheet resistance of MoS$_2$, rather than that of the specific contact resistivity within the field-emission region. For the single-layer MoS$_2$ case, we achieved the contact resistance of ~1 kΩ μm at $T = 100$ K, which is the lowest value reported by metal evaporations on MoS$_2$. Our experimental findings were corroborated by *ab initio* DFT and DFT-MGF calculations. Comparison with theory showed that the In metal weakly interacts with MoS$_2$, or an ideal In/MoS$_2$ contact is a relatively FLP-free interface with an Ohmic *n*-type band alignment. Importantly, we showed that the band upshift across the In/MoS$_2$ vdW gap is abrupt and rigid, while the below-CBM states that contribute to the Ohmic transport can be attributed to MIGS.

## Methods

**TEM imaging** A cross-sectional TEM sample was prepared by the dual-beam focused ion beam (Helios, FEI). HR-FE-TEM (JEM-2100F, JEOL) was used for the TEM imaging at 200 kV.

**Raman spectrum** The Raman measurements were performed in a backscattering geometry at room temperature. An incident laser light with a wavelength of 514.5 nm was focused on the sample surface through an optical microscope objective lens (100×/0.9 NA). An excitation laser power was maintained less than 0.4 mW to avoid any laser-induced heating effects. Scattered light from the sample was dispersed through a monochromator with a 1200 grooves/mm grating and was collected using a thermoelectrically cooled charge-coupled device detector. For mapping measurements, Raman spectra were taken at the step of 0.5 μm over the area of 15 × 15 μm$^2$.

**Electrical measurements** To apply an electric field to the $MoS_2$ channel, the highly *p*-doped Si substrate was biased by a back-gate voltage ($V_G$). All measurements were performed in a cryostat (PPMS, Quantum Design, Inc.) with a base *T* of 2.4 K. The $I$-$V_{SD}$ curves were measured by a DC bias voltage source (Yokogawa 7651) combined with a current pre-amplifier (DL 1211). The two-probe and four-probe conductance measurements for $1L$-$MoS_2$ device were performed by a dc measurement setup. The two-probe and four-probe conductance measurements for $6L$-$MoS_2$ device were performed by using a standard lock-in amplifier (SR830) with current preamplifier, where excitation voltage and output frequency were 30 mV and 77.77 Hz, respectively.

**DFT and DFT-MGF calculations**. We adopted the VASP software[27] to carry out DFT geometry optimizations for the metal/$MoS_2$ slab models within the Grimme DFT-D3 scheme[28] until the Hellmann-Feynman forces were below 0.02 eV· $Å^{-1}$. We then calculated their electronic structure using the SIESTA software[29] within the vdw-DF2 functional.[30] To avoid artificial interactions with neighbour unit cell images within the periodic boundary condition, we inserted a vacuum space of more than 15 Å along the slab-normal direction. For the quantum transport calculations, we constructed the junction models starting from the In/$Mo_2$ models described above (Fig. 5c; see also Supplementary Fig. S9) and carried out DFT-MGF calculations using the TranSIESTA code.[31] We confirmed that the computational results are not modified by adopting a different exchange-correlation functional (Supplementary Table 1). More details of DFT and DFT-MGF calculations can be found in Supplementary Fig. S10.

## Acknowledgments

We acknowledge Prof. E. Hwang and Dr. J.-S. Park for fruitful discussion. This work was supported by the Korea Research Institute of Standards and Science (KRISS-GP2018-003), part of the Basic Science Research Program through the National Research Foundation of Korea (NRF) (Grant Nos. 2018R1A2A1A05078440, SRC2016R1A5A1008184, 2016R1A2B4008525 and 2019R1A2C1003366). This work was also partly supported by the Korea-Hungary Joint Laboratory Program for Nanosciences through the National Research Council of Science and Technology. Growth of hexagonal boron nitride crystals was supported by the Elemental Strategy Initiative conducted by the MEXT, Japan and JSPS KAKENHI Grant Numbers JP15K21722. Y.-H.K. and T.H.K were supported by the NRF grants (Basic Research Program 2017R1A2B3009872, Nano-Material Technology Program 2016M3A7B4024133, Global Frontier Program 2013M3A6B1078881, and Basic Research Lab Program 2017R1A4A1015400). Computational resources were provided by the KISTI Supercomputing Center (KSC-2018-C2-0032).

## Author contributions

M.-H.B. and J.-J.K conceived the research project and Y.-H.K provided the theoretical interpretation. K.W. and T.T grew the bulk hBN. D.-H.C and B.-K.K fabricated the devices. D.-H.C performed the TEM analysis. B.-K.K performed the electrical measurements and analysed the data with M.-H.B. H.K. and H.R. performed the Raman spectroscopy. T.-H.K performed DFT and DFT-MGF calculations. All authors wrote the manuscript.

**Figure captions**

**Figure 1| Characterization and electrical properties of In/MoS₂ devices. a,** In (Au) evaporation in a vacuum chamber with a liquid-nitrogen cooled sample holder. **b,** Top and bottom: cross-sectional TEM image of the In/MoS₂ and Au/MoS₂ interfaces, respectively. Scale bar: 5 nm. **c,** Optical image of a 6L-MoS₂/hBN device with multiple In contacts. Scale bar: 5 μm. **d,** Conductance ($G$) as a function of the back-gate voltage ($V_G$) at various temperatures.

**Figure 2| Contact and sheet resistance of an In/6L-MoS₂ device. a,** $R_c W$ as a function of carrier density ($n_e$) of In/6L-MoS₂ device at various temperatures (scattered solid squares) with previous works (solid curves: graphene(Gr)/4L-MoS₂, opened diamonds: Au/4L-MoS₂, opened pentagons: Au/In/few L-MoS₂). **b,** $R_c W$ and $R_{sh}$ as a function of $T$ at $n_e = 3.4 \times 10^{12}$ cm⁻².

**Figure 3| Contact resistance vs. sheet resistance and $R_c W$ of 1L-MoS₂. a,** Minimum $R_c W$ as a function of sheet resistance ($R_{sh}$) of In/6L-MoS₂ device at various temperatures (scattered solid squares) with previous works (opened triangles: graphene(Gr)/4L-MoS₂, opened diamonds: Au/4L-MoS₂, opened pentagons: Au/In/few (1) L-MoS₂). The upper and lower dotted lines were obtained with $\rho_c = 5 \times 10^{-6}$ Ω cm² and $5 \times 10^{-7}$ Ω cm², respectively. **b,** Four-probe $G$–$V_G$ curves at various temperatures obtained from the $L = 1$ μm channel, indicated by a dashed box in the inset of **b**. The inset of **b**: AFM image of a 1L-MoS₂ device with four In electrodes. The region outlined by a solid white line indicates the MoS₂ flake. Scale bar: 5 μm. **c,** $R_c W$–$T$ curves corresponding to various $V_{G-th}$ conditions.

**Figure 4| Electron accumulation-type contact and Raman mapping of In/1L- and In/2L-MoS₂. a,b,** Optical images of 1L- and 2L-MoS₂ (dashed black-boxed region) on SiO₂ substrates, respectively. White boxed regions: 5 nm-thick In-deposited regions. Scale bar: 5 μm. **c,d,** $A_{1g}$ energy ($\omega$) maps for 1L- and 2L-MoS₂, respectively. **e,f,** Full-width at half-maximum ($\Gamma$) maps of $A_{1g}$ for 1L- and 2L-MoS₂, respectively.

**Figure 5| DFT and DFT-based matrix Green's function calculations at In(Au)/MoS₂ interface. a,** DFT-optimized geometries and interfacial charge transfers obtained from the In/MoS₂ (top) and Au/MoS₂ (bottom) interface models. Red and blue colors represent the charge accumulation and depletion regions, respectively. The isosurface level is 0.0007 e/Å³. **b,** For the In/MoS₂ interface model (top panel; the red dashed box indicates the supercell atomic structure for DFT calculations), we show the projected DOS (bottom panel) of the top S $3p$ states (bottom left panel), Mo $4d$ states (bottom central panel), and bottom S $3p$ states (bottom right panel). Shown together is the bottommost In $5p_z$ states (filled curves). The arrows indicate the prominent MIGS. **c,** Half of the In/MoS₂/In junction model adopted for DFT-MGF calculations (top panel) and the calculated local DOS (bottom panel). The red (A), blue (B), and green (C) upper triangles indicate the In-contact, interface, and intrinsic channel MoS₂ regions, respectively. CBM and VBM indicate the conduction band minimum and valance band maximum, respectively. **d,** Schematics of the In and MoS₂ band levels before contact. **e** and **f,** Schematics of the previously proposed band alignments for vdW contacts based on the band bending. **g,** Schematic band alignment derived from our DFT-MGF calculations (Fig. 5c). Here, the red lines represent MIGS.

**Figures**

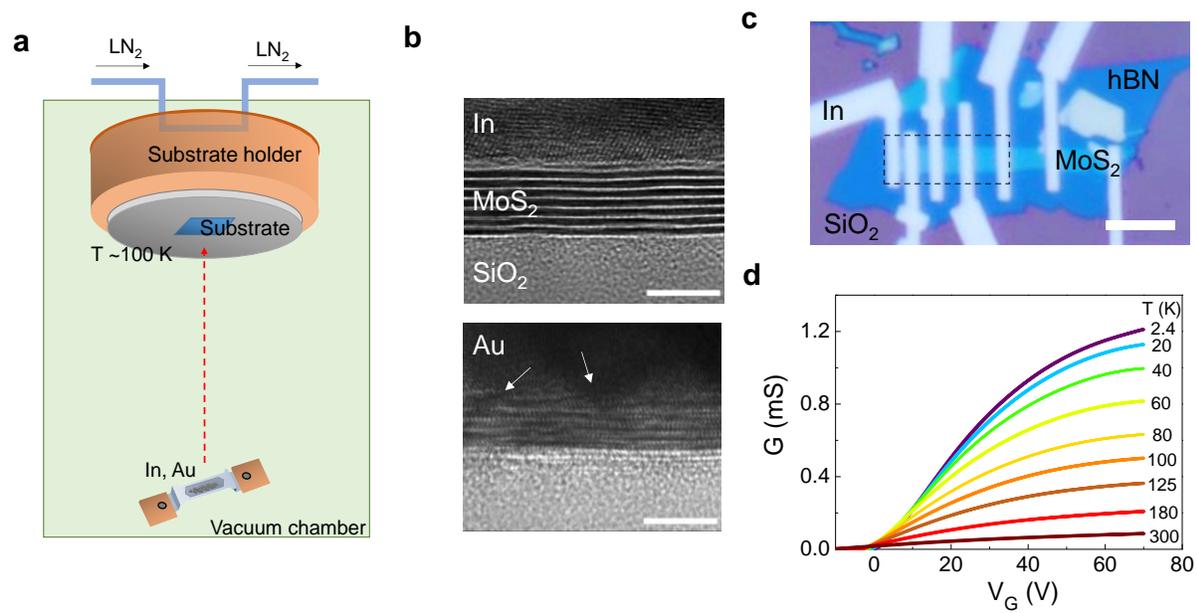

**a**

LN₂    LN₂

Substrate holder

Substrate

T ~100 K

In, Au

Vacuum chamber

**b**

In

MoS₂

SiO₂

Au

**c**

In    hBN

MoS₂

SiO₂

**d**

T (K)
2.4
20
40
60
80
100
125
180
300

**Figure 1**

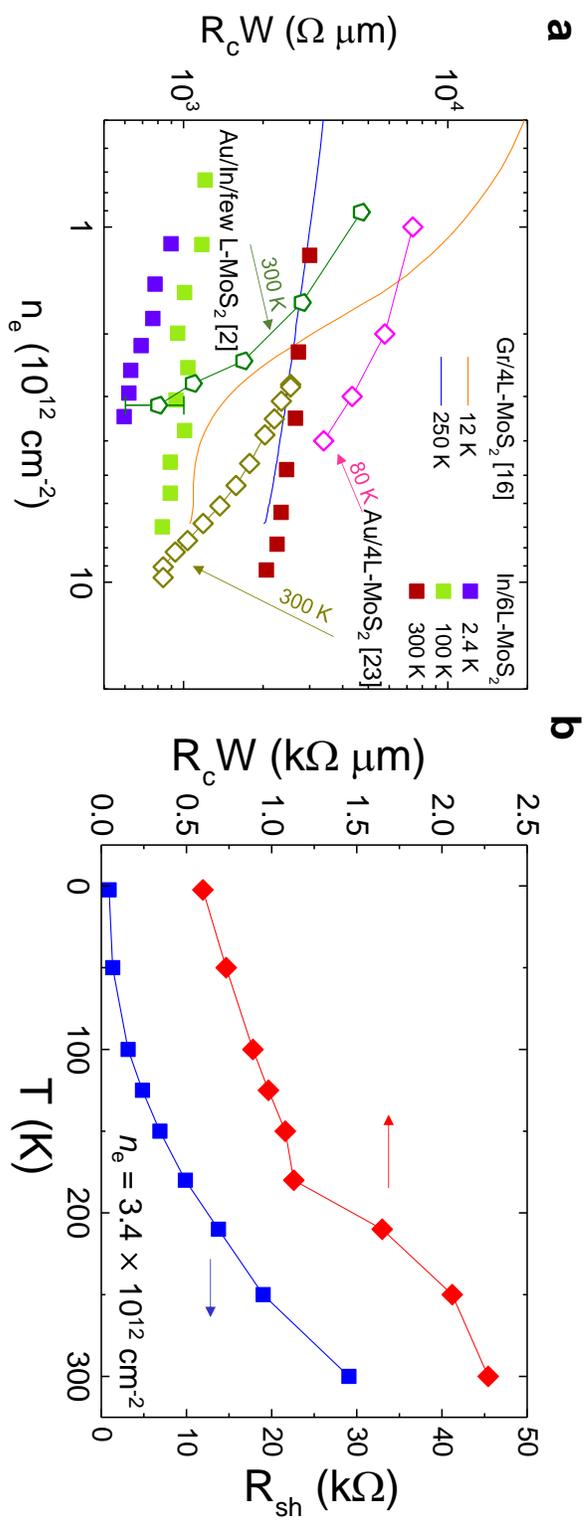

**Figure 2**

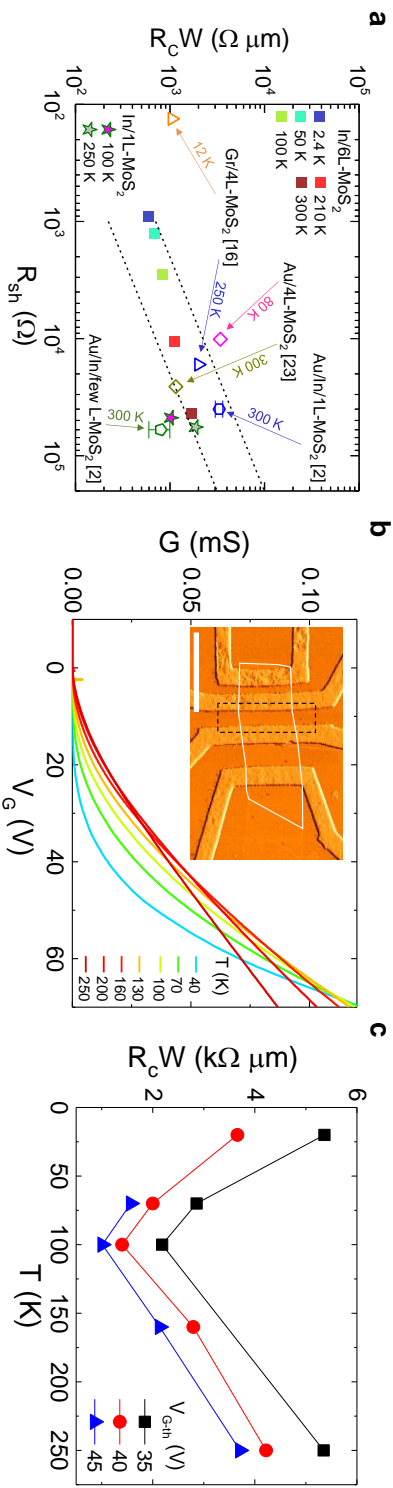

**Figure 3**

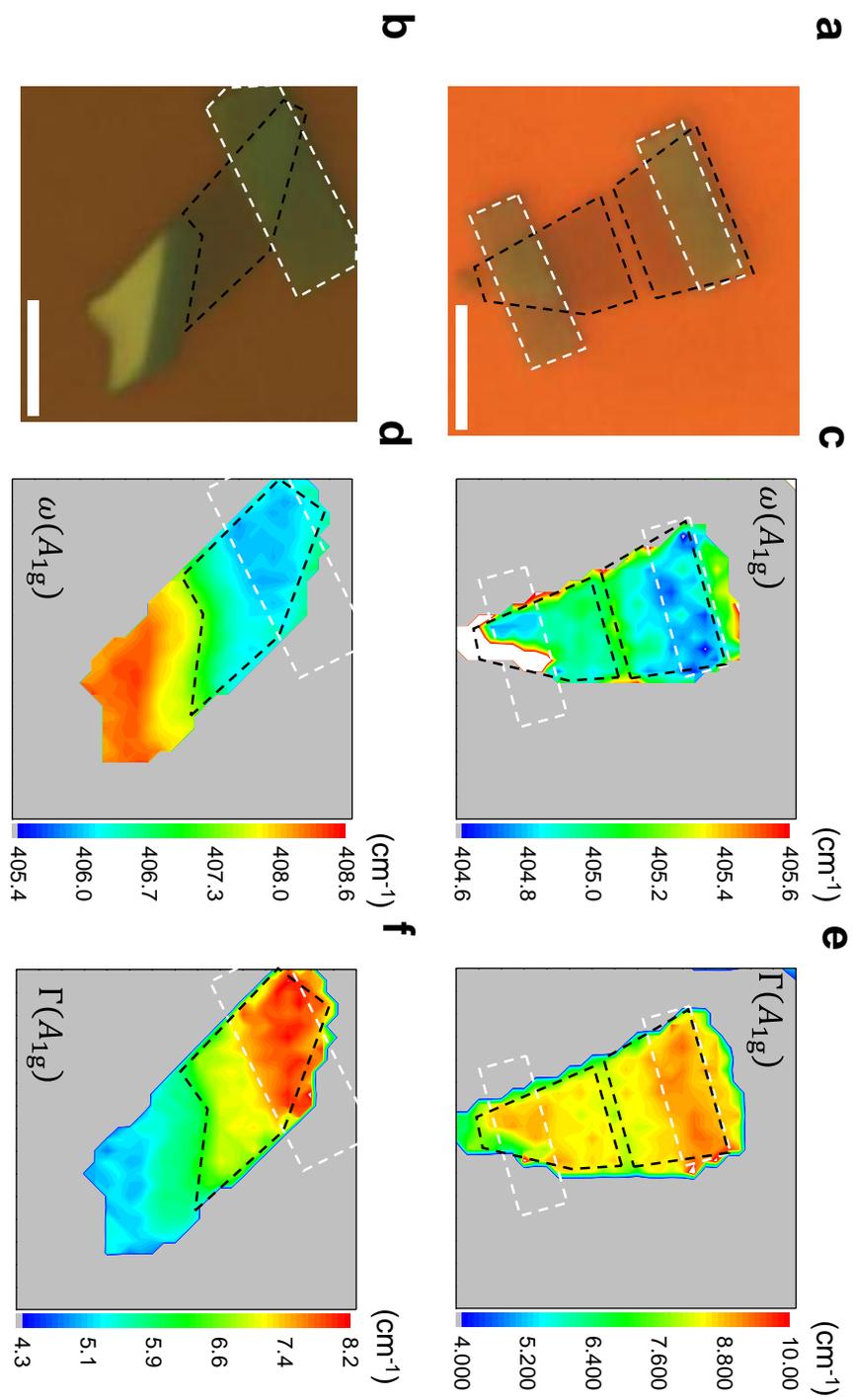

**Figure 4**

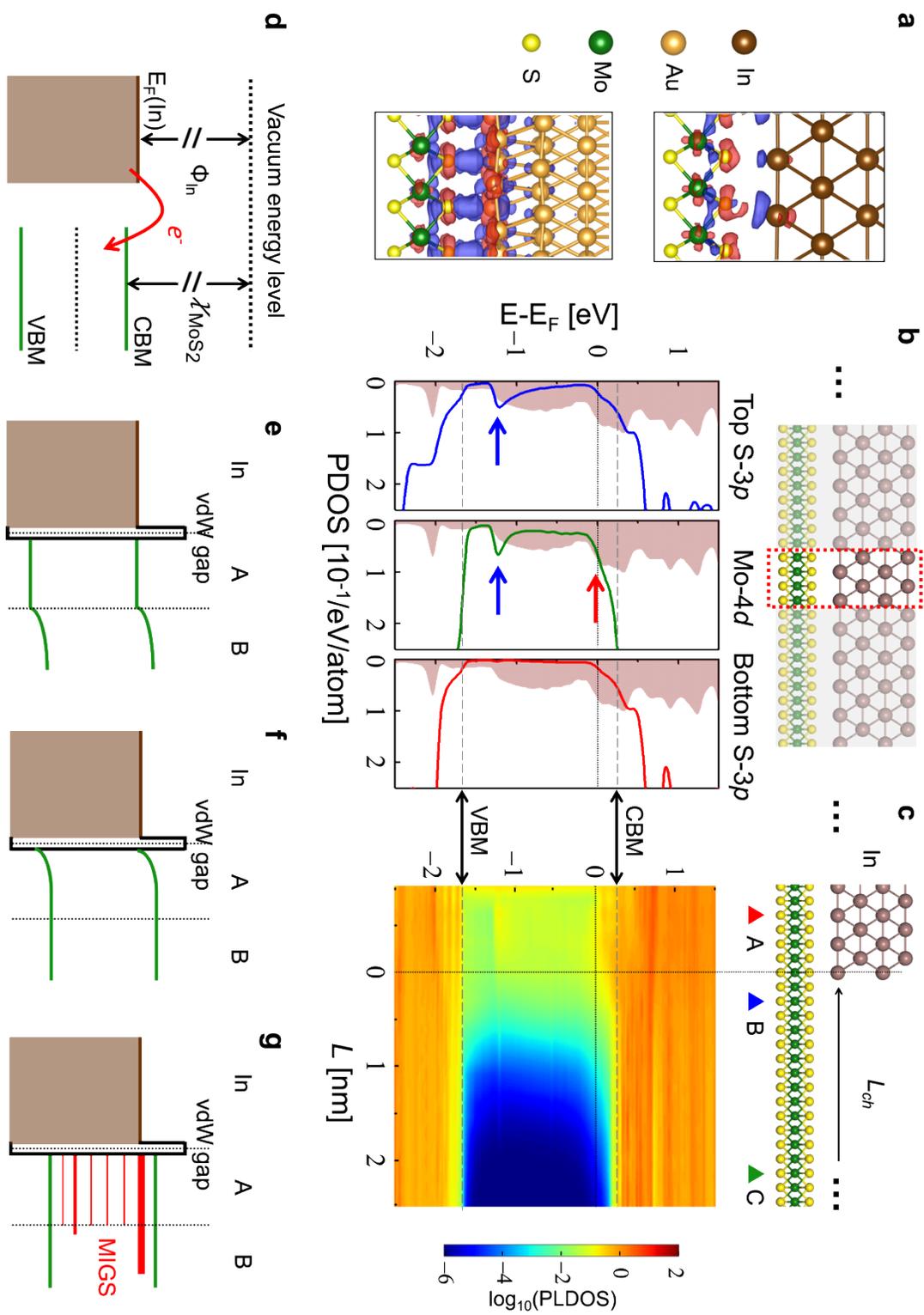

**Figure 5**

# Supplementary Information for

## Genuine Ohmic van der Waals contact between indium and MoS$_2$

Bum-Kyu Kim[1†], Dong-Hwan Choi[1,2†], Tae-Hyung Kim[3†], Hanul Kim[2], Kenji Watanabe[4], Takashi Taniguchi[4], Heesuk Rho[2], Yong-Hoon Kim[3*], Ju-Jin Kim[2*], and Myung-Ho Bae[1,5*]

[1]Korea Research Institute of Standards and Science, Daejeon 34113, Republic of Korea

[2]Department of Physics, Research Institute of Physics and Chemistry, Chonbuk National University, Jeonju 54896, Republic of Korea

[3]School of Electrical Engineering, Korea Advanced Institute of Science and Technology, 291 Daehak-ro, Yuseong-gu, Daejeon 34141, Republic of Korea

[4]National Institute for Materials Science, 1-1 Namiki, Tsukuba 305-0044, Japan

[5]Department of Nano Science, University of Science and Technology, Daejeon, 34113, Republic of Korea

[†]These authors contributed equally to this work.

*e-mail: y.h.kim@kaist.ac.kr; jujinkim@chonbuk.ac.kr; mhbae@kriss.re.kr

**Section 1. Six-layer (6L) MoS₂ device**

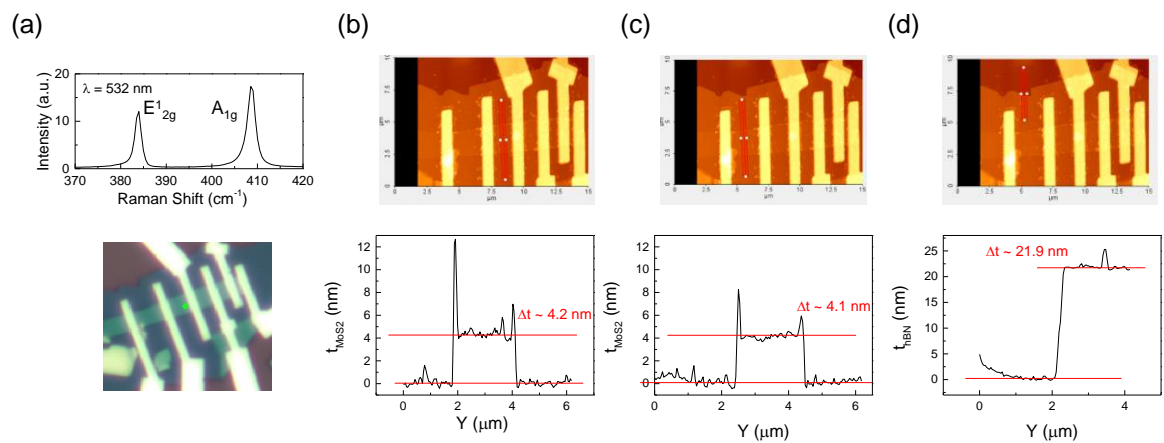

Figure S1. (a) Raman spectrum of 6L-MoS₂ device (upper panel), which was taken at the green spot in a photo-image of the lower panel. (b)-(d) upper panels: AFM images of the device. Lower panels: thickness profile along the red boxes in the corresponding upper panels ((b),(c): MoS₂, (d): hBN).

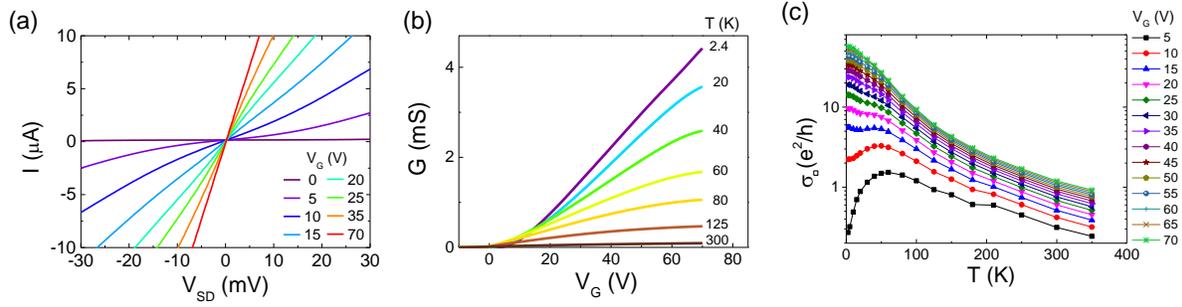

Figure S2. (a) Current ($I$) as a function of the source–drain voltage ($V_{SD}$) for various $V_G$ at $T$ = 2.4 K. (b) 4-probe $G$-$V_G$ curves of 6L-MoS$_2$ device for various temperatures. (c) Normalized conductance as a function $T$ for various $V_G$. For $V_G > 15$ V corresponding to the crossover temperature between the metal and insulator characters, the down-turn curvature for $T < 70$ K was disappeared.

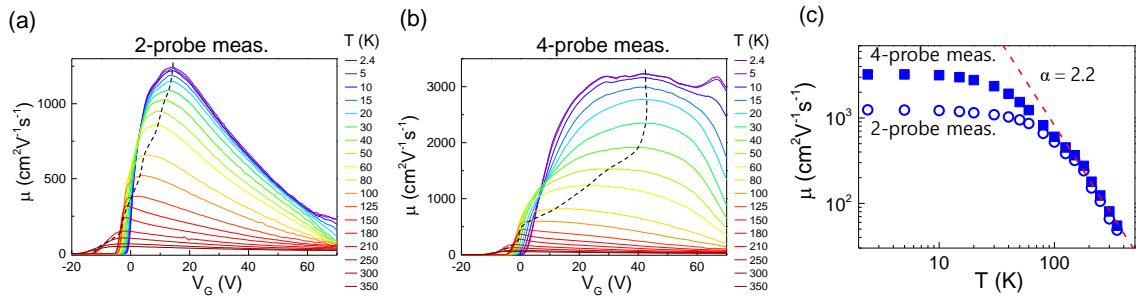

Figure S3. Mobility (μ) as a function of $V_G$ at various temperatures for (a) two-probe and (b) four-probe measurements. The dashed lines traces the local maximum mobility for varying $T$. The mobility was obtained by a relation of $\mu = \frac{L}{W}\frac{1}{C_G}\frac{dG}{dV_G}$, where $L$ and $W$ are channel length and width, respectively, and $C_G \left(= \left(C_{SiO_2}^{-1} + C_{hBN}^{-1}\right)^{-1}\right)$ is the back-gate capacitance. (c) μ as a function of temperature ($T$) on a log scale, as obtained from 2-probe (open circles) and 4-probe (closed squares) measurements. The dashed line is a fitting line with a relation of $\mu(T) \propto T^{-\alpha}$ ($\alpha$ = 2.2).

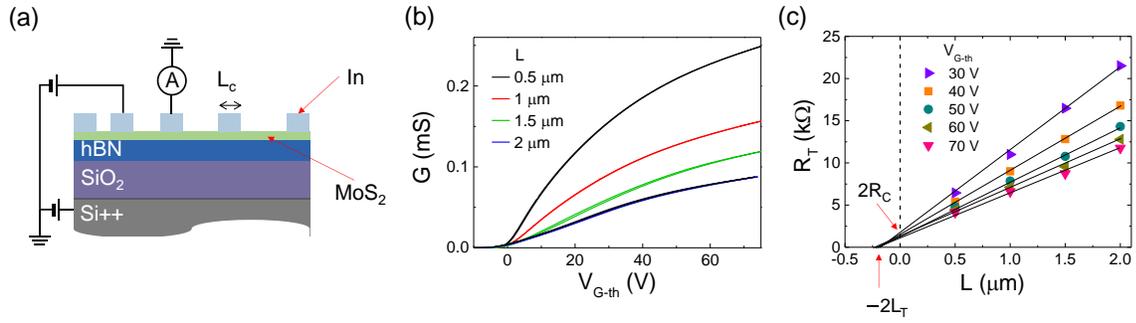

Figure S4. (a) Schematic of 6L-MoS$_2$ device for the transfer length method (TLM). Here, $L_c$ (= 1 μm) is the length of contact electrode. The measurement scheme shows the two-probe measurement for the $L_2$ = 1 μm channel. (b) Conductance as a function of $V_{G-th}$ for the four channels at $T$ = 210 K. (c) Total resistance ($R_T$) as a function channel length $L$ for representative $V_{G-th}$ values at $T$ = 210 K. $R_c$ and $L_T$ are the contact resistance and transfer length, respectively. These are indicated by arrows at $V_{G-th}$ = 30 V, as an example. $L_T$ was estimated as ∼0.1 μm.

**Section 2. Single-layer (1L) MoS₂ device**

(a)

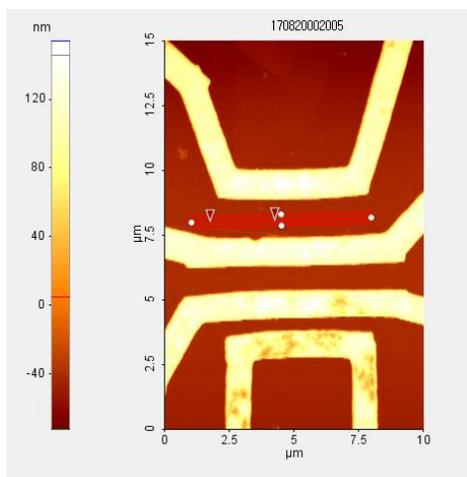

(b)

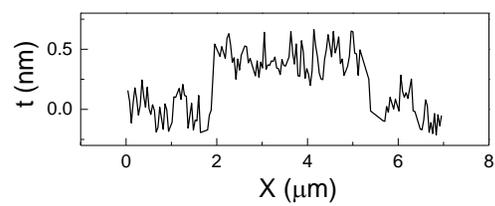

Figure S5. (a) AFM image of 1L-MoS₂ device. (b) Thickness profile along the red line in (a).

**Section 3. 1L-MoS₂ to study the doping effect by In contacts**

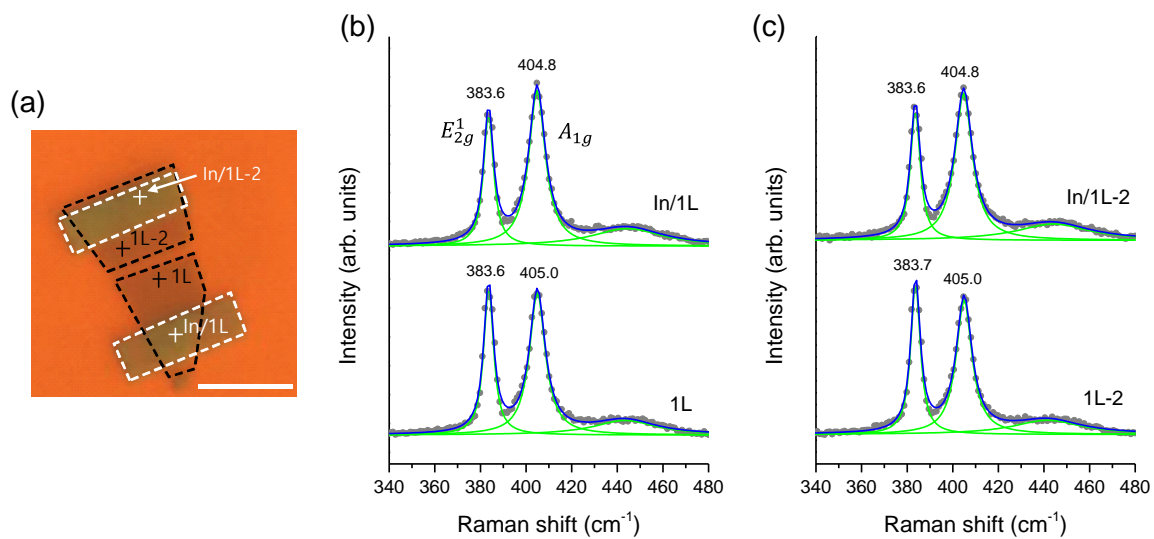

Figure S6. (a) Optical photo-image of a 1L-MoS₂ (outlines: dashed black lines) on SiO₂ with indium contacts (dashed white lines). (b) and (c) Scattered points: Raman spectra obtained at white (In/1L) and black (1L) cross marks in (a). Green and blue curves are Lorentzian-fit results to estimate the energy and width of phonon modes.

**Section 4. First-principles density functional theory (DFT) and matrix Green's function (MGF) calculations**

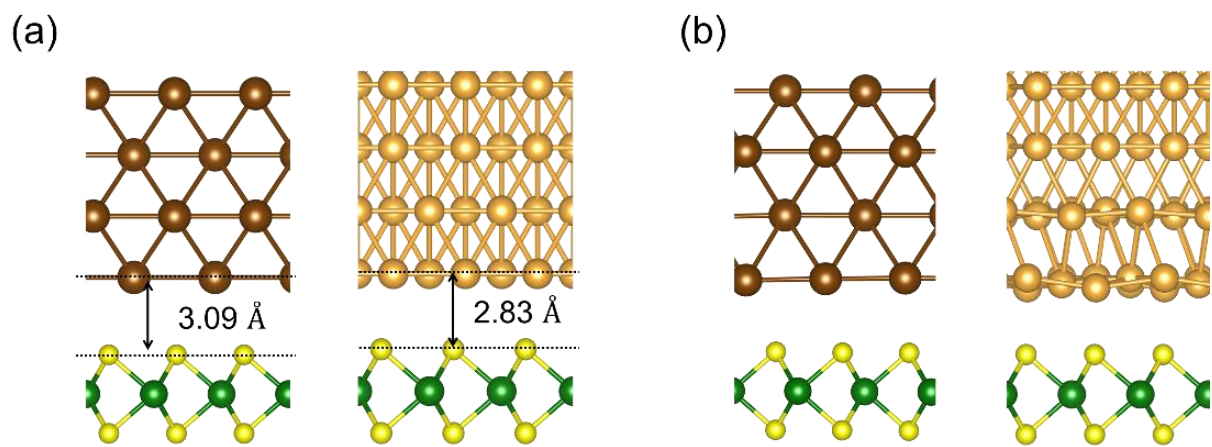

Figure S7. (a) Vertical interface model of In/MoS$_2$ (left panel) and Au/MoS$_2$ (right panel) before geometry optimization. Here, the MoS$_2$-metal distance was determined with Grimme DFT-D3 method. (b) Optimized structures for In/MoS$_2$ (left panel) and Au/MoS$_2$ (right panel), respectively.

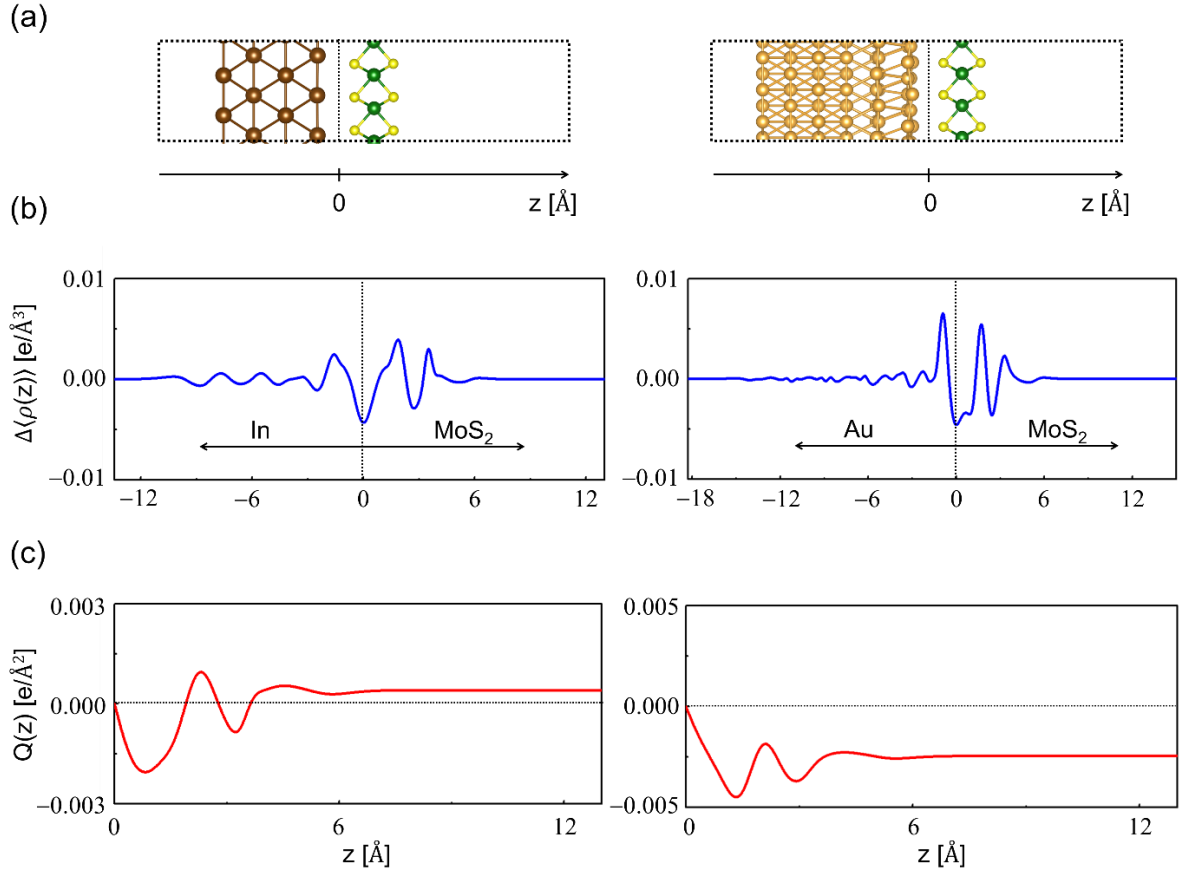

Figure S8. (a) Vertical In(Au)/MoS$_2$ model (from left to right). (b) Plane averaged charge density difference ($\Delta\langle\rho\rangle = \langle\rho_{Total}\rangle - \langle\rho_{MoS_2}\rangle - \langle\rho_{Metal}\rangle$) for the vertical In(Au)/MoS$_2$ model (from left to right). We chose a reference position (z=0) at the minimum $\Delta\langle\rho\rangle$ position at the metal-S interface. (c) Net charge transfer $Q$(z) from In(Au) to the MoS$_2$, which are calculated according to $Q(z) = \int_0^{z_{vacuum}} \Delta\langle\rho\left(z^{'}\right)\rangle dz^{'}$. The positive (negative) $Q$(z) indicates an electron transfer from metal to MoS$_2$ (from MoS$_2$ to metal).

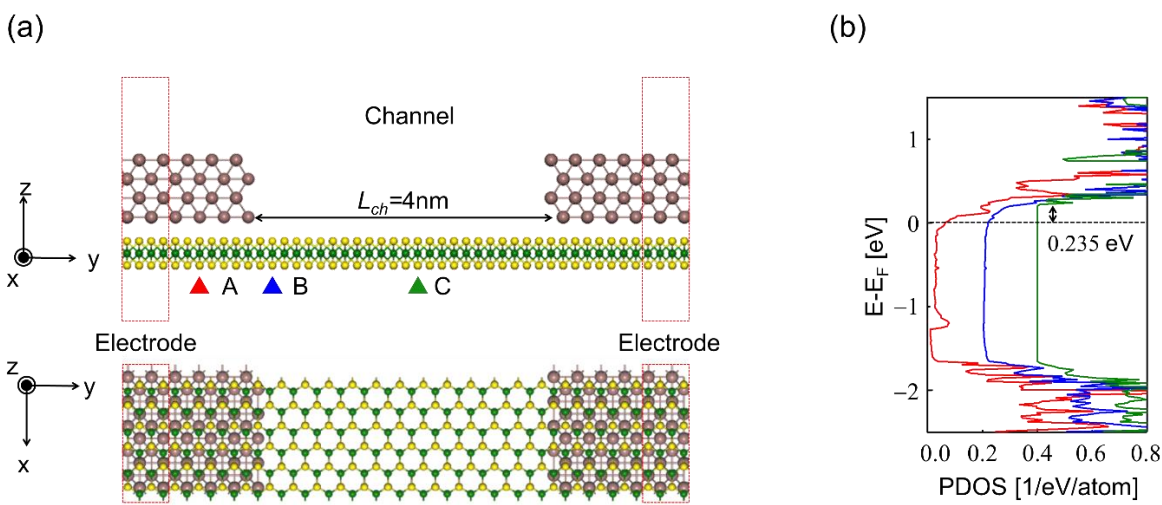

Figure S9. (a) Models for the MGF calculations for the In/MoS$_2$/In junction. Here, red-dashed rectangular boxes indicate the semi-infinitely repeated left and right electrode parts within quantum transport calculations. The A, B, and C upper triangles indicate the electrode contact-region, transition-region, and intrinsic channel-region MoS$_2$ positions, respectively. (b) MGF-calculated Mo-4$d$ projected local density of states (PLDOS). The red, blue, and green lines correspond to the Mo-4$d$ PLDOS at A, B, and C positions, respectively. The zero of energy is shifted by 0.2 eV and 0.4 eV for the PLDOS at B and C positions, respectively.

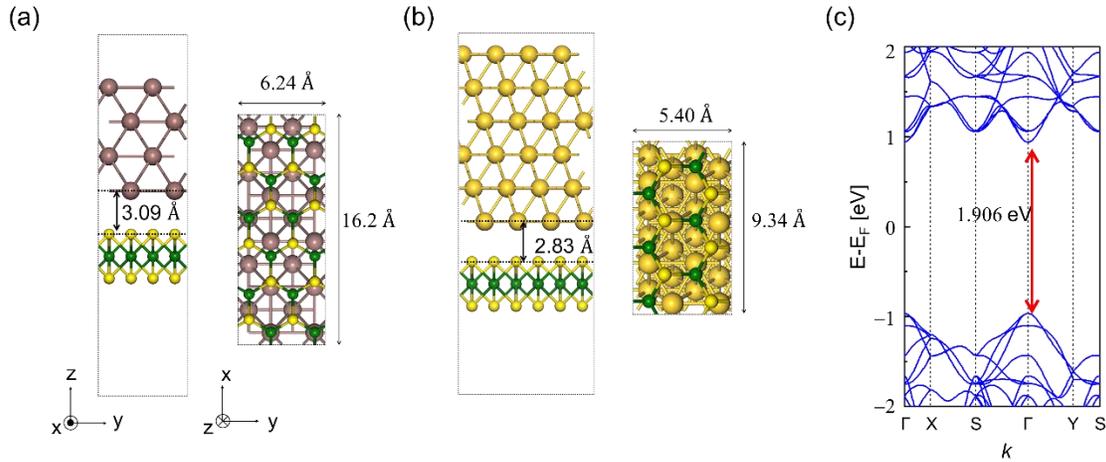

Figure S10. **DFT and DFT-MGF calculation details.** DFT calculation models: (a) In/MoS$_2$ and (b) Au/MoS$_2$. (c) Band structure for pristine single layer MoS$_2$ in the Au/MoS$_2$ cell box. Each structure of Au, In, MoS$_2$ was obtained from local density approximation (LDA)[1] calculation, since each structure is homogenous. Then, we made vertical interface models and the junction model for quantum transport calculation. Here, we adopted the lattice parameters for the hetero-structure models as the lattice parameters of MoS$_2$. The lattice mismatch for In-MoS$_2$ and Au-MoS$_2$ is 0.31% and 6.58%, respectively. Then, a MoS$_2$-metal distance was determined with Grimme DFT-D3 scheme [2] to capture van der Waals (vdW) interaction. After modelling, we adopted the vdW-DF2 functional [3] to capture non-local vdW interaction in the electronic structures. All electronic structures were obtained from vdW-DF2. Next, we confirmed that the pristine single layer MoS$_2$ has a direct band gap (~1.9 eV) with vdW-DF2, which is well matched previous study [4]. During DFT and MGF calculations, the atomic cores were replaced by norm-conserving nonlocal pseudopotentials of the Troullier-Martins type. Then, we adopted the double ζ-plus-polarization-level numerical atomic orbital basis sets with 30 meV energy cut-off for MoS$_2$. Meanwhile, we chose the optimized basis cut-off values for each metal in order to correctly describe metal work function. Then, we chose a $k$-points mesh for In/MoS$_2$, Au/MoS$_2$, and In/MoS$_2$/In as

| Functional | Metal | WF [eV] | $\chi_{MoS_2}$ [eV] | $E_g$ [eV] |
|---|---|---|---|---|
| **vdW-DF2** | In | 4.121 | 4.189 | 1.906 |
| | Au | 5.137 | 4.189 | 1.906 |
| **LDA** | In | 4.112 | 4.196 | 1.941 |
| | Au | 5.130 | 40196 | 1.941 |

| | Metal | $Q$ [e/Å$^2$] | $\Delta V_{H|MoS_2-M}$ [eV] | BH$_C$ [eV] |
|---|---|---|---|---|
| **vdW-DF2** | In | $3.61 \times 10^{-4}$ | 0.422 | 0.235 |
| | Au | $-1.97 \times 10^{-3}$ | -0.411 | 0.537 |
| **LDA** | In | $7.19 \times 10^{-4}$ | 0.525 | 0.334 |
| | Au | $-1.27 \times 10^{-3}$ | -0.219 | 0.630 |

Table S1. Comparison with vdw-DF2 and LDA results for the vertical In(Au)/MoS$_2$ interface models. Here, we used an ideal van der Waals contact models as shown in Fig. S9a and 10a, b. We compare the metal work functions (WF), electron affinity for MoS$_2$ ($\chi_{MoS_2}$), band gap for pristine single layer MoS$_2$ ($E_g$), net charge transfer from metal to MoS$_2$ ($Q$(z)), electrostatic potential difference between MoS$_2$ and metal ($\Delta V_{H|MoS_2-M}$), and barrier height for region C (BH$_C$) in the junction model (Fig. S9). Specifically, we calculated the amount of net charge transfer, $Q$(z) with same criterion as mentioned in Fig. S8. Next, the plane averaged electrostatic potential difference between MoS$_2$ and metal was calculated as following equation, $\Delta V_{H|MoS_2-M} = \Delta V_{H|MoS_2} - \Delta V_{H|Metal}$, where the plane averaged electrostatic potential was calculated as $\Delta V_H = V_{H|Total} - (V_{H|Metal} + V_{H|MoS_2})$. Based on such information, we have found that the calculation results are almost similar regardless functional.

**References for Supplementary Information**